\documentclass[twocolumn,prl,aps,superscriptaddress]{revtex4}
\usepackage{latexsym}
\usepackage[dvips]{graphicx}

\begin{document}

\title{Observation of a new phase transition between fully and partially
polarized quantum Hall states with charge and spin gaps at $\nu = \frac{2}{3}$}
\author{N.
Freytag}\email[corresponding author: ]{freytag@polycnrs-gre.fr}
\affiliation{Grenoble High Magnetic Field Laboratory, MPI-FKF and CNRS, B.P.166,
F-38042 Grenoble Cedex 9, France}
\author{Y. Tokunaga}
\affiliation{Grenoble High Magnetic Field Laboratory, MPI-FKF and CNRS, B.P.166,
F-38042 Grenoble Cedex 9, France}
\author{M. Horvati\'{c}}
\affiliation{Grenoble High Magnetic Field Laboratory, MPI-FKF and CNRS, B.P.166,
F-38042 Grenoble Cedex 9, France}
\author{C. Berthier}
\affiliation{Grenoble High Magnetic Field Laboratory, MPI-FKF and CNRS, B.P.166,
F-38042 Grenoble Cedex 9, France}
\affiliation{Laboratoire de Spectrom\'{e}trie Physique, Universit\'e J. Fourier,
B.P.  87, F-38402 St.  Martin d'H\`{e}res, France}
\author{M. Shayegan}
\affiliation{Department of Electrical Engineering, Princeton
University, Princeton N.J.  08544}
\author{L. P. L\'evy}
\affiliation{Grenoble High Magnetic Field Laboratory, MPI-FKF and CNRS, B.P.166,
F-38042 Grenoble Cedex 9, France}
\affiliation{Institut Universitaire de France and Universit\'e J. Fourier, B.P.  41,
F-38402 St.  Martin d'H\`{e}res, France}
\date{\today}

\begin{abstract}

  The average electron spin-polarization $\cal P$ of two-dimensional
  electron gas confined in $\rm GaAs/GaAlAs$ multiple quantum-wells was
  measured by nuclear magnetic resonance (NMR) near the fractional quantum
  Hall state with filling factor $\nu=\frac{2}{3}$.  Above this
  filling factor (${ \frac{2}{3}} \leq \nu < 0.85$), a strong
  depolarization is observed corresponding to two spin flips per
  additional flux quantum.  The most remarkable behavior of the
  polarization is observed at $\nu ={ \frac{2}{3}}$, where a quantum
  phase transition from a partially polarized (${\cal P} \approx {
    \frac{3}{4}}$) to a fully polarized (${\cal P} = 1$) state can be
  driven by increasing the ratio between the Zeeman and the Coulomb
  energy above a critical value $\eta_{\rm c} = \frac{\Delta_{\rm
      Z}}{\Delta_{\rm C}} = 0.0185$.

\end{abstract}

\maketitle




The integer quantum Hall effect (QHE) occurs when the Landau level
(LL) filling factor $\nu$ is an integer $p$ and the lowest $p$ LL's
are filled \cite{QHE}. However, if the Zeeman splitting $\Delta _{{\rm
    Z}}=|g|\mu _{{\rm B}}B$ is increased with respect to the LL
spacing $\hbar \omega_{\rm c}$ ($g=0.44$ is the electronic Land\'{e}
factor in GaAs, $\mu _{{\rm B}}$ the Bohr magneton, $B$ the magnetic
field and $\omega_{\rm c}$ the cyclotron frequency), crossings of LL's
with different spin states occur when $k \cdot \hbar \omega_{\rm c} =
\Delta_{\rm Z}$. If the level crossing occurs for LL's at the Fermi
energy, the spin configuration of the electrons changes \cite{Koch93}.
The fractional QHE (FQHE) occurs when $ \nu = \frac{p}{2mp \pm 1}$,
where $m =1,2$ and can be visualized with the composite fermion (CF)
model \cite{CF}. In this picture FQHE corresponds to integer QHE of
CF's, where the CF-LL spacing is now determined by the Coulomb energy
$\Delta _{{\rm C}}=\frac{1}{4\pi \epsilon _{0}\epsilon _{{\rm r}}}
\frac{e}{\ell _{B}}\propto \sqrt{B_{\perp}}$ ($\epsilon _{0} \epsilon
_{{\rm r}}$ is the permittivity GaAs, $ \ell _{B}=\sqrt{\hbar /
  eB_{\perp}}$ is the magnetic length, the average distance of two
electrons). Hence similar spin transitions do also occur in the FQHE.
Since the relevant energy scales for CF's are the Zeeman and the
Coulomb energies, the spin configuration of a 2DEG is mainly driven by
the exchange part of the Coulomb energy ($\propto \sqrt{B_{\perp}}$)
at low $B$, whereas at large fields, the Zeeman energy ($\propto B$)
favors spin-polarized states.  For example, $\nu = \frac{2}{3}$
corresponds to two filled CF-LL's: only two states are possible,
${\cal P}=0$ at low $B$ and ${\cal P}=1$ in the opposite limit. The
transition between these states are controlled by the ratio $\eta
=\frac{\Delta _{{\rm Z}}}{\Delta _{{\rm C}} }$ at a given $\nu $. This
ratio can be changed by varying the electron density ($n$) or by
rotating the angle ($\theta $) between the normal to the 2DEG and the
applied magnetic field, keeping the perpendicular field $B_{\perp}$
constant (tilted field technique).

In this Letter, we report an NMR observation of a phase transition
between a new, partially polarized QH phase and the fully polarized
high field phase at $\nu = \frac{2}{3}$. The partially polarized phase
has charge and spin gaps and is completely unexpected by theory.
Furthermore, this is the first observation of a phase transition
between QH states {\em without} the gap going through a minimum at the
transition point.




In this study, two GaAs/GaAlAs multiple (100) quantum well (QW)
samples M280 (M242) with electron densities $n=8.5\cdot 10^{10}$
($1.4\cdot 10^{11}$) $ {\rm cm^{-2}}$ and mobilities $\mu _{0}=7\cdot
10^{5}$ ($3\cdot 10^{5}$) $ \frac{\mbox{cm}^{2}}{\mbox{Vs}}$ have been
used \cite{Samples}. The $w=$ 30 (25) nm wide GaAs QWs are separated
by 250 (185) nm thick ${\rm Al_{0.1}Ga_{0.9}As}$ (${\rm
  Al_{0.3}Ga_{0.7}As}$) barriers with Si $\delta $-doping near their centers.
Two pieces of each sample, $\approx 25\mbox{ mm}^{2}$ in size, were
placed in the center of a radio-frequency (RF) coil, so that 200 QWs
effectively contribute to the nuclear magnetic resonance (NMR) signal.
At higher temperatures ($1.5\mbox{ K}\leq T\leq 10\mbox{ K}$), the NMR
experiment was performed in a variable temperature insert as a function
of $T$, $B$, and $\theta $. At lower temperatures ($40\mbox{ mK}\leq
T\leq 1.5 \mbox{ K}$), the RF-coils were mounted at fixed $\theta $
into the mixing chamber of a dilution refrigerator.

NMR is a very sensitive {\em direct} measurement of the spin
polarization of 2D-electrons in the QWs, since the Fermi contact
interaction ${\cal H}=\frac{8\pi}{3}|\gamma_e|\gamma_n
\hbar^2\sum_{ij} {\cal S}_i \cdot {\cal I}_j\delta({\vec r}_i-{\vec
  R}_j)$ ($\gamma$ is the gyromagnetic ratio) between itinerant
electron spins ${\cal S}_i$ at position ${\vec r}_i$ and nuclear spins
${\cal I}_j$ at ${\vec R}_j$ shifts the resonance frequency of
$^{71}$Ga nuclei in the QWs by a magnetic hyperfine shift $K_{{\rm
    S}}$ proportional to ${\cal P}$ \cite{1nu2paper,NMR1, NMR2}.  The
NMR signal from barriers (without electrons) remains unshifted and is
used as a reference.  If the electronic motion is rapid with respect
to the NMR timescale ($\approx 1$-$10$ $\mu$s), the nuclei in QWs see
an average value $\langle {\cal S} \rangle $. Hence, the average
spin-polarization is inferred from the hyperfine shift from ${\cal
  P}(\nu ,T,\theta )=\frac{K_{{\rm S}}(\nu ,T,\theta )}{K_{{\rm
      S}}({\cal P}=1)}$, where $K_{{\rm S}}({\cal P}=1)$ is the
maximum shift measured in each sample for high Zeeman energy.  Values
of $K_{{\rm S}}({\cal P}=1)$\ for our two samples obey the empirical
definition of the ``coupling constant'' $A_{c}=\frac{w}{n} K_{{\rm
    S}}({\cal P}=1)\approx 4.5\cdot 10^{-13}~$cm$^{3}$/s, determined
previously by optically pumped NMR \cite{OPNMR}.  In contrast to other
quantitative measurements of ${\cal P}$ which use optical techniques
\cite{Kukushkin}, standard NMR is a quasi-equilibrium probe of the
2DEG polarization and does not affect the electron system.

At high $T$, where $K_{\rm S}$ is small, we take advantage of the much
shorter nuclear spin-lattice relaxation times $T_{1}$ of Ga nuclei in
the QWs compared to those in the barriers in order to distinguish
their contributions to the signal \cite {1nu2paper}.  An NMR
pulse-sequence first destroys the nuclear magnetization $ M_{0}$ in
the whole sample.  After a recovery time $t_{\rm R}$ of the order of
$T_1$ of the QWs nuclei, few nuclei in the barriers have recovered and
the NMR signal is dominated by the nuclei in the wells.  The spectrum
is obtained either by free induction decay after a $\frac{\pi }{2}$
pulse or by a $\frac{\pi }{2}$--${\pi}$ spin-echo technique.  

\begin{figure}[tb]
\centering
  \includegraphics*[width=80mm]{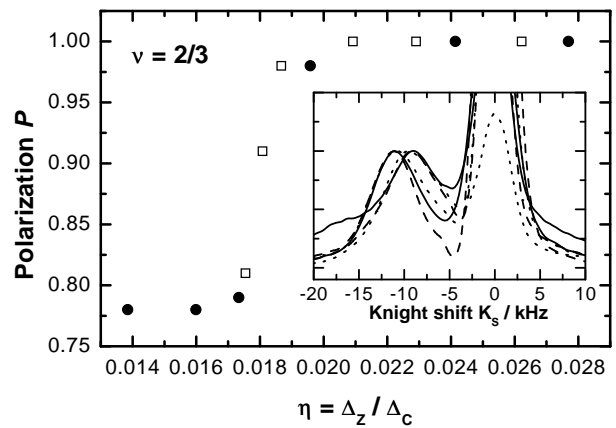}
\caption{The phase transition of the $\nu = \frac{2}{3}$ state as revealed 
  by the dependence of the spin polarization on the ratio of Zeeman and
  Coulomb energies $\protect\eta = \frac{\Delta_{{\rm
        Z}}}{\Delta_{{\rm C}}}$ for Sample M280 ($\bullet$) and M242
  ($\Box$) at temperatures below $100$ mK. $\eta$ was varied by
  changing the angle $\theta$ between the magnetic field direction and
  the normal to the 2D plane. The inset shows the NMR spectra
  corresponding to the points for M280, where the QWs signal amplitude
  is renormalized.}
\label{PvsEta}
\end{figure}

At low temperature, where $K_{{\rm S}}$ is sufficiently large to
distinguish both QWs and barriers' line positions in a single spectrum
(see inset Fig.~\ref{PvsEta}), a simple read-out sequence with
small ``tip-angle'' pulses spaced by a long waiting time \cite{FID} is
preferred, in order to avoid heating the sample and spurious effects
on the electron system.



\begin{figure}[tb]
\centering
  \includegraphics*[width=80mm]{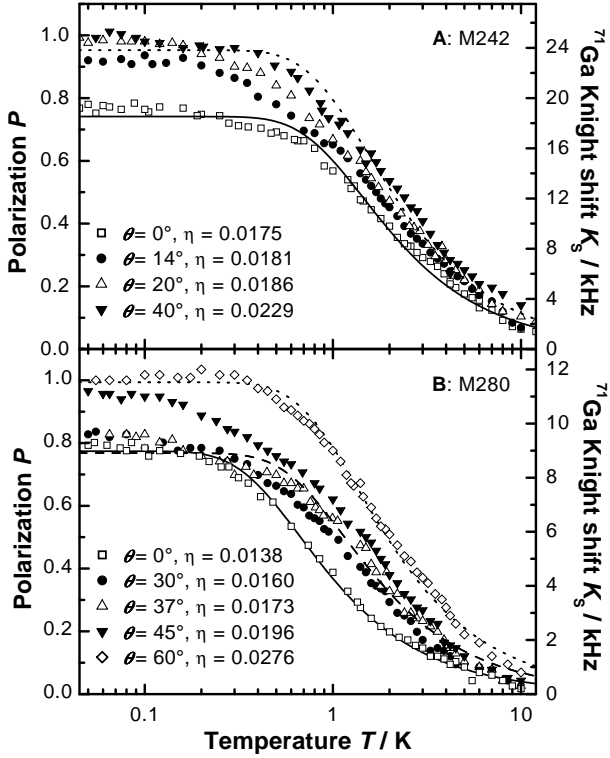}
\caption{The temperature dependence of the spin polarization at 
  $\nu = \frac{2}{3}$ for Sample M242 (A) and M280 (B) at various tilt
  angles $\protect\theta $. The magnetic field is given by $5.325/\cos
  \protect\theta $ T for M280 and $8.55/\cos \protect \theta $ T for
  M242. The lines are least square fits of a two-level model to the
  data (M280: $0^{\circ}$, $37^{\circ}$ and $60^{\circ}$, M242:
  $0^{\circ}$ and $40^{\circ}$).}
\label{PvsT}
\end{figure}

The dependence of the electron spin polarization on $\eta
=\frac{\Delta _{ {\rm Z}}}{\Delta _{{\rm C}}}$ at the exact filling
factor $\nu ={\ \frac{2}{3}} $ is shown in Fig.~\ref{PvsEta} for both
samples M280 and M242, as obtained from the low temperature saturation
values of $\cal P$ at various tilt angles $\theta$ (see Fig.
\ref{PvsT}). We observe a Zeeman energy induced phase transition from
a {\em partially} polarized ground-state (GS) with an average ${\cal \ 
  P}\approx {\ \frac{3}{4}}$ to a {\em fully} polarized GS above the
{\em same} critical value $\eta _{{\rm c}}\approx 0.018$ for both
samples (the total magnetic fields at this critical point are $
B_{c}^{M280}=6.7~$T and $B_{c}^{M242}=8.9~$T).  The transition from
one GS to the other is very sharp, within a change of $\eta $ of about
3\%. Such a sharp jump in spin-polarization suggests a first-order
phase transition. The precision of this measurement can be seen from
the inset to Fig.~\ref{PvsEta} where the spectra for M280 are plotted.
The error on the line's position is less than 0.1 kHz.

A full polarization at large $\eta $ qualitatively agrees with
numerical calculations on small systems~\cite{QHE,Chakraborty1} where
the fully spin-polarized GS at $\nu ={\frac{2}{3}}$ has been found to
have the lowest energy for large Zeeman energy. In this limit, there
is a one to one correspondence between $\nu = \frac{1}{3}$ and
$\frac{2}{3}$.  The $\frac{2}{3}$ state can be regarded as the
particle-hole conjugated state of $\frac{1}{3}$ within the lowest LL.
When Zeeman energy is reduced the symmetry between these states is
broken.  Whereas the $\nu = \frac{1}{3}$ state is still fully
polarized, $\frac{2}{3}$ is an unpolarized state. Also within the
composite fermion (CF) model at $\nu =\frac{2}{3}$ only ${\cal P}=0$
and ${\cal P}=1$ are possible.  The $\nu =\frac{2}{3}$ state is built
out of the fully polarized or unpolarized $\nu=2$ state by attaching
two flux quanta, pointing anti-parallel to the external magnetic
field, to each electron and projecting the wavefunction on the lowest
LL~\cite{CFspin}. Within this picture a partial polarization ${\cal
  P}\approx {\frac{3}{4}}$ {\em cannot} be constructed from filled
CF-LL's, but have to be considered as an inhomogeneous mixture of
${\cal P}=1$ and ${\cal P}=0$ states.  Similarly, this value of the
polarization cannot be reached by exact diagonalization on a small
system unless at least $8\cdot k$ electrons are included.  This
exceeds the limit of existing state-of-the-art calculations already
for $k>2$. Recently Apalkov {\it et\ al.}\ found a transition upon
increasing $\eta $ from an unpolarized to half polarized, and to fully
polarized GS within a 12 electron calculation \cite{Apalkov}.

There are other pieces of evidence for QH phases at $\nu =\frac{2}{3}$
with $ {\cal P}\neq 1$. Using optical methods, Kukushkin {\it et al.}\ 
\cite {Kukushkin} have observed partially polarized GS at $\nu
=\frac{2}{3}$ with $ {\cal P}=\frac{1}{2}$ at $\eta \approx 0.008$. In
their measurement the polarization rises smoothly from ${\cal P}=0$ at
$\eta <0.007$ to ${\cal P}=1 $ for $\eta >0.01$ with a narrow plateau
at half polarization. Nevertheless, transport measurements on electron
doped samples at $\nu =\frac{2}{3}$ are displaying a deep minimum of
the thermal activation gap at $\eta \approx 0.01$.  This indirect
measurement was interpreted as a transition from unpolarized to fully
polarized states ${\cal P}=0\rightarrow {\cal P}=1$
\cite{Transport2nu3}. In our samples, where $\eta(\theta = 0^{\circ
  })\geq 0.0138$, the transport gap at $\nu =\frac{2}{3}$
monotonically increases with increasing Zeeman energy, even across the
transition at $\eta _{c}$ \cite{Sorin}.

\begin{figure}[tb]
\centering
  \includegraphics*[width=70mm]{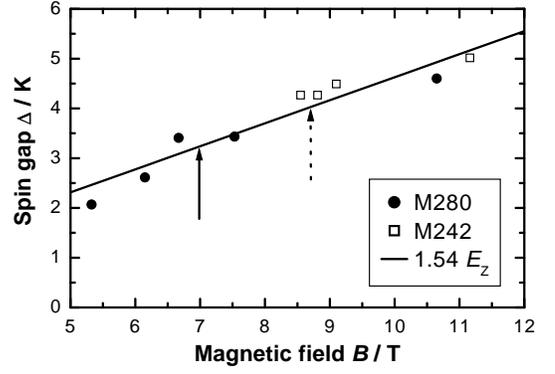}
\caption{The magnetic field dependence of the spin gap for Sample M280 
  ($\bullet$) and M242 ($\Box$) extracted from the temperature
  dependence of the spin polarization shown in Fig. \ref{PvsT}. The solid
  line corresponds to $1.54 \cdot E_{\rm Z}$ and the arrows indicate
  the transition points for both samples.}
\label{gapvsB}
\end{figure}

The temperature dependence of the polarization at $\nu ={\frac{2}{3}}$
for both samples at various tilt-angles is presented in
Fig.~\ref{PvsT}. The lines are fits to the data using a two-level model: 
assuming the partition function factorizes
into a spin-independent and a spin-dependent part, the temperature
dependence of the polarization is $ {\cal P}(B,T)={\cal
  P}_{T=0}(B)\cdot \tanh \left( \frac{\Delta }{4k_{{\rm B} }T}\right)
$, where the gap $\Delta =g^{\ast }\mu _{{\rm B}}B$ ($g^{\ast }$ is
the enhanced $g$-factor) is the splitting between the spin-up and
spin-down components. Away from the transition ($\eta \not\approx \eta
_{c}$), this fit describes well the data (M280: $\eta =0.0138$,
$0.0276$ and M242: $\eta =0.0229$).
But close to the transition ($\eta \approx \eta_c$), a two-level model
fails to describe the temperature dependence of ${\cal P}$: the
saturation value $ {\cal P}_{T=0}$ of ${\cal P}$ is only reached at
very low $T$ , while ${\cal P}$ is reduced at intermediate
temperatures.  This may be a consequence of a coexistence region
between two phases with different polarizations.  This model defines a
thermally activated spin-gap $\Delta $ which varies smoothly across
the transition at $\eta _{c}$. Such a dependence rules out a simple
``level crossing transition'' between two magnetic GSs, where $\Delta
$ should collapse at the transition, and suggest a first-order
transition.  We extracted $\Delta$ even in the case where the two
level model fails by determining the temperature where half of the
saturation value is reached: $\Delta \approx 2.2 \; T({\cal P}={\cal
  P}_{T=0} / 2)$. These gaps are plotted in Fig. \ref{gapvsB}. They
correspond to $\Delta = (1.54\pm 0.04)\,\Delta _{{\rm Z}}$ and yield a
50\% exchange enhancement of the Zeeman splitting $\Delta _{{\rm Z}}$.

\begin{figure}[tb]
\centering
  \includegraphics*[width=80mm]{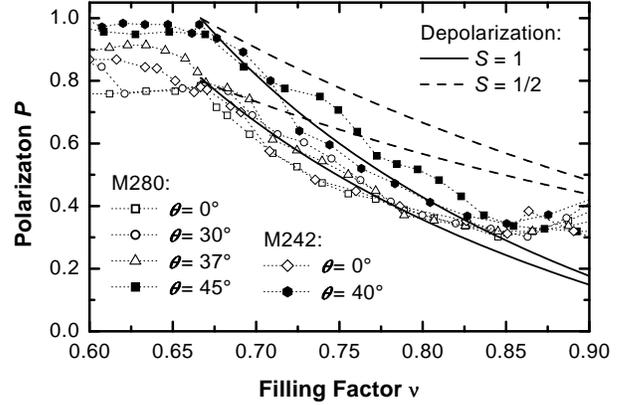}
\caption{The filling factor dependence of the polarization for samples M280
  and M242 at $T = 100$ mK and various tilt angles $\protect \theta$.
  Solid symbols correspond to full polarization at $\protect\nu ={\ 
    \frac{2}{3}}$ (M280: $\protect\theta=45^{\circ}$ and M242:
  $\protect\theta =40^{\circ}$) and open symbols (M280:
  $\protect\theta=0^{\circ}$, $ 30^{\circ} $, $37^{\circ}$ and M242:
  $\protect\theta=0^{\circ}$) to partial polarization; thin dotted
  lines are guide to the eye. The solid (dashed) lines simulate a
  depolarization of {\em two} (one) spin flips, $\Delta m_S = -2$,
  ($\Delta m_S=-1$) per removed flux quantum.}
\label{PvsNu}
\end{figure}

In Fig.~\ref{PvsNu}, the filling factor dependence of the spin
polarization is shown for both samples and various tilt angles. The
most surprising observation is the strong depolarization measured
above filling factor $\nu ={\frac{2}{3}}$, whether the polarization at
$\nu ={\frac{2}{3}}$ is full or partial. The data below and above the
transition represented in Fig.~\ref{PvsNu} as open and filled symbols
respectively, superpose on one line for both samples, M280 and M242.
This spin depolarization can be well reproduced over a wide range of
$\nu $ if we assume {\em two} spin flips per removed flux quantum.
This is shown by solid red lines in Fig.~\ref{PvsNu}. For comparison,
a depolarization corresponding to a single spin flip per flux quantum
is also shown in dotted red lines, clearly out of the observed
behavior. The two experimental depolarization lines converge around
$\nu \approx 0.85$ towards a polarization close to $\frac{1}{3}$.

For filling factors lower than $\nu ={\frac{2}{3}}$, the behavior of
${\cal P}(\nu )$\ is radically different; the polarization remains
{\em constant} for $\nu < \frac{2}{3}$, for both full and partial
polarization. The exception to this behavior are only the data sets
(M280: $\theta =37^{\circ }$, M242: $\theta =0^{\circ }$) {\em at} the
transition $\eta \approx \eta _{c}$, where polarization increases from
the partial towards the full polarization when $\nu $\ is reduced
below ${\frac{2}{3}}$.

Before discussing these data, we recall that the number of orbital
states available in one LL is equal to the number of magnetic field
flux quanta $D= \frac{\phi }{\phi _{0}}=\frac{e{\cal A}}{\hbar }\cdot
B$ (${\cal A}$ is the sample area, $\phi $ the flux through the
sample, and $\phi _{0}$ the flux quantum). When $\frac{2}{3}$ of the
available states are occupied by electrons ($\nu \equiv n{\cal
  A}/D={\frac{2}{3}}$), the system condenses in the $\frac{2}{3}$-FQH
state, a state with precisely 3 flux quanta per 2 electrons. If the
magnetic field is slightly decreased or increased, the filling factor
will be slightly higher or lower than $\nu =\frac{2}{3}$, but the
lowest energy GS will remain to be the same $\frac{2}{3}$-FQH state,
now containing the quasi-holes or quasi-particles (QP)
corresponding to ``extra'' or ``missing'' flux quanta~\cite{Girvin}.
The QP excitations at $\nu =\frac{2}{3}$\ carry a charge of
$\frac{e}{3}$ and there are $\frac{3}{2}\phi_0$ per electron. Hence
the QP corresponds to $\frac{1}{3}\cdot \frac{3}{2}\phi_0 =
\frac{1}{2}\phi_0$, meaning that per added $\phi_0$ {\em two} QPs are
added. This can easily provide spin ${\cal S}=1$ or $0$ (and never
${\cal S}=\frac{1}{2} $).

The same conclusion also holds in a naive CF picture, where the $\nu =
\frac{2}{3}$ state is obtained by a mapping from the $\nu = 2$ state.
Increasing the field, {\em i.e.}\ injecting additional flux quanta,
increases the degeneracy of the LLs. The quasi-holes added in this way
are placed into the same, {\em spin-aligned} LLs as the CFs which
does not modify the polarization of the system of CFs, as is
experimentally observed {\it below} $\nu ={\frac{2}{3}}$.  Decreasing
the field is equivalent to adding QPs to the system of CFs.  Per
missing $\phi_0$ there is one QP per CF-LL and at $\nu = \frac{2}{3}$
there are two occupied CF-LLs. If both QPs are occupying {\em
  spin-reversed} CF-LLs, this provides precisely the depolarization
with two reversed spins per additional $\phi_0$, as is observed {\it
  above} $\nu =\frac{2}{3}$,




There is no satisfactory understanding of the FQHE in the low magnetic
field limit. In particular, since filled CF-Landau levels can only
produce fully polarized or unpolarized states at $\nu=\frac{2}{3}$ ,
the nature of a partially polarized state at this filling factor is
open.  Among the possible states, it is natural to consider some of
the other known spin-liquid states which have been considered in the
literature such as the valence-bond GS of the 2D Shastry-Sutherland
model~\cite{Shastry}. Other well-known spin-gapped states are
commensurate spin-density waves.

An alternative viewpoint is to consider partially polarized states as
inhomogeneous mixture of fully polarized and unpolarized regions \cite
{oneline}. The GS could also be described as isolated singlets forming
a commensurate modulated structure through the sample, in many ways
similar to valence-bond commensurate spin-liquid GSs. Naturally, the
real challenge is to determine among all the possible GSs, which one
has the lowest energy.

In conclusion, this NMR study has revealed a new phase transition
between FQH ground-states at $\nu ={\frac{2}{3}}$\ with different
spin-polarizations.  Considering the sharpness of the jump in
polarization as well as the thermal activation close to $\eta _{c}$,
the transition appears to be first order.  Whereas the quasi-hole
excitations for $\nu < \frac{2}{3}$ are spin-aligned, the
quasi-particle excitations for $\nu > \frac{2}{3}$ are found to be
spin-reversed, independent from the transition, leading to a rapid
depolarization above $\nu = \frac{2}{3}$.




We gratefully acknowledge enlightening discussions with R. Morf.




\begin{references}
\bibitem{oneline}  A single NMR line is observed: if spin-mixture is
present, it is not static.
\end{references}
\end{document}